# Recovering Latent Structure in Massive Datasets: A PCA Study of 10 Billion and 1 Trillion Observations

By: Dr. Mike Crowhurst*
Graduate Student, Dr. Bing Zhang Department of Statistics
University of Kentucky

***Abstract***

This study investigated the behavior of Principal Component Analysis (PCA) when applied to datasets with extremely large numbers of observations. Although statistical theory suggests that sampling error diminishes and sample estimates converge toward their population values as sample size increases, relatively little empirical evidence exists regarding the behavior of PCA at scales measured in billions or trillions of observations. Three datasets were analyzed: a 10-billion observation random dataset, a 1-trillion observation random dataset, and a 10-billion observation engineered dataset designed to contain three latent factors.

Results showed that the PCA solutions obtained from the 10BillionRandom and 1TrillionRandom datasets were nearly identical, indicating substantial stability of PCA at extremely large sample sizes. In contrast, the engineered dataset produced three dominant principal components that accounted for 99.996% of the total standardized variance and successfully recovered the intended latent-factor structure. These findings suggest that PCA solutions converge rapidly at very large sample sizes and suggest that PCA solutions may reach practical convergence well before sample sizes reach the trillions. These findings have implications for large-scale applications in fields such as remote sensing, digital mapping, environmental modeling, and other domains where datasets routinely contain millions or billions of observations.

***Introduction***

In a previous article, (Crowhurst, 2026) we explored the use of a GPU-accelerated multivariate statistics platform and the use of Principal Component Analysis (PCA) in the analysis of a very large dataset. The dataset in that study was a randomly generated dataset with 10-billion observations. PCA was used to examine the structure of the dataset and to evaluate the feasibility of conducting multivariate analyses at previously impractical scales. Although PCA identified a set of principal components, the analysis revealed no dominant low-dimensional structure. Variance was distributed across multiple principal components, and no small subset of components accounted for a substantial proportion of the total variance. This pattern is consistent with what would be expected from randomly generated data.

The earlier study generated two important questions. First, does PCA exhibit practical convergence when sample size increases from billions to trillions of observations generated from the same underlying process? Classical statistical theory suggests that estimates should stabilize as sample size increases, but relatively little empirical evidence exists at these scales. Second, can PCA

*AI-based tools were used for language refinement and for assistance in developing Python code used in the simulations. All methodological decisions, results, and interpretations are the responsibility of the author.*

successfully recover known latent structure when applied to a dataset containing billions of observations that has been intentionally engineered to contain a predefined component structure?

The purpose of the present study was to re-use the mechanism used in the first paper first on a new dataset with 10-billion randomly generated observations (10BillionRandom), second on a dataset with 1-trillion randomly generated observations (1TrillionRandom), and finally on a dataset with 10-billion observations that was specifically designed to produce three principal components (10BillionEngineered). The random datasets were expected to exhibit little or no meaningful low-dimensional structure, whereas the engineered dataset was expected to produce three dominant principal components corresponding to the latent factors used in data generation.

All analyses were conducted using a GPU-accelerated workstation; details of the computing environment are provided in the Methods section.

### ***Background and Motivation***

The volume of data available for scientific, commercial, and governmental applications has increased dramatically over the past several decades. Early discussions of "Big Data" emphasized the increasing volume, velocity, and variety of information generated by modern information systems (Laney, 2001). As storage capacities, computational resources, and network infrastructure have improved, datasets containing billions of observations have become increasingly common in fields such as finance, healthcare, social media analytics, e-commerce, sensor networks, and scientific computing. Manyika, et al. (2011) argued that the ability to collect and analyze large datasets has become a major driver of innovation, productivity, and decision-making across numerous industries.

The rapid growth of data has also created significant challenges for statistical analysis. Traditional statistical methods were often developed and evaluated using datasets containing hundreds or, at most, thousands of observations. Many were developed before computers and relied on hand calculations. As dataset sizes increased to millions and billions of records, researchers began to investigate the computational and methodological implications of large-scale analysis. Fan, Han, and Liu (2014) noted that Big Data introduces challenges related not only to computation and storage but also to the interpretation and validation of statistical results. As sample sizes increase, it becomes important to determine whether statistical procedures continue to produce stable and meaningful results.

PCA is one of the most widely used techniques for dimensionality reduction, feature extraction, and latent-structure discovery (Jolliffe, 2002). PCA transforms a set of correlated variables into a smaller number of orthogonal components that explain the majority of variation present in a dataset. Because of its ability to summarize complex multivariate relationships, PCA has been applied extensively in the physical sciences, social sciences, engineering, finance, and machine learning.

Despite the widespread use of PCA, relatively little research has examined its behavior when applied to datasets containing billions or trillions of observations. Classical statistical theory suggests that sampling error diminishes and sample estimates converge toward their population values as sample size increases. As the number of observations grows, sample means, variances, covariances, and correlations are expected to converge toward their corresponding population values. In finite populations, sampling error approaches zero as sample size approaches population size. Similarly, for data generated from an underlying probability distribution, statistical estimates are expected to become increasingly stable as the number of observations becomes very large. Although this convergence is well understood theoretically, relatively little empirical research has examined the practical behavior of multivariate techniques such as PCA when applied to datasets containing billions or trillions of observations. An important question is whether PCA solutions obtained from datasets containing billions of observations differ meaningfully from those obtained using datasets containing trillions of observations when both datasets are generated from the same underlying process. From a practical perspective, determining the point at which additional observations produce negligible changes in PCA results may be as important as understanding the large sample behavior of statistical estimators.

A second motivation for this study concerns the recovery of known latent structure. PCA is often used in exploratory analyses where the underlying structure of the data is unknown. A stronger test of PCA occurs when the true latent structure is known in advance. If variables are intentionally generated from a small number of latent factors, PCA should recover a corresponding set of dominant principal components. Examining whether this recovery occurs in extremely large datasets provides additional evidence regarding the validity and stability of PCA-based analyses at scale.

The present study was motivated by the intersection of two questions. The first concerns statistical convergence: at what point does increasing sample size cease to produce meaningful changes in the results of a classical multivariate procedure such as PCA? The second concerns latent-structure recovery: can PCA identify known component structure when that structure is embedded within datasets containing billions of observations? By comparing PCA solutions obtained from 10-billion and 1-trillion observation random datasets, and by examining an engineered dataset with a known three-factor structure, this study provides an empirical examination of both convergence and latent-structure recovery at previously unexplored scales.

## ***Methods***

### *Study Design*

Three datasets were analyzed. The first dataset contained 10 billion randomly generated observations (10BillionRandom). The second dataset contained 1 trillion randomly generated observations (1TrillionRandom). The third dataset contained 10 billion observations generated from a predefined latent-factor model designed to produce three principal components (10BillionEngineered).

The two random datasets were used to evaluate the stability of PCA as sample size increased from 10 billion to 1 trillion observations. The engineered dataset was used to evaluate the ability of PCA to recover a known latent-factor structure at extreme scale.

### *Computing Environment*

All analyses were conducted on a single accelerator-rich computing node equipped with five NVIDIA graphics processing units (GPUs). The system operated under Microsoft Windows 11 and software development was performed using C++ with CUDA version 12.8 for GPU acceleration (NVIDIA Corporation, 2026). The computing node contained 32 GB of system memory and a 4 TB solid-state drive for local storage of intermediate results and output files.

*Data Generation*

The two random datasets (10BillionRandom and 1TrillionRandom) were generated using the variable definitions shown in Table 1. Each observation contained ten variables derived from three independently generated random variables. The resulting datasets were intended to provide a baseline condition in which no explicit latent-factor structure was imposed.

Table 1 – Variables in original CSV file.

| Column | Description | Formula |
|---|---|---|
| 1 | Random Number Between 3 and 8 | int B = randBetween(3, 8); |
| 2 | Random Number Between 1 and 10 | int C = randBetween(1, 10); |
| 3 | Random Number Between 1 and 100 | int D = randBetween(1, 100); |
| 4 | Log of C / Log of B * 100 | double E = int(log(C) / log(B) * 100); |
| 5 | Log of D / Log of B * 10000 | double F = round((log(D) / log(B)) * 10000); |
| 6 | Integer of the Absolute Value of the Cosine of C * 100 | double G = int(abs(cos(C)) * 100); |
| 7 | Integer of the Absolute Value of the Sine of D * 100 | double H = int(abs(sin(D)) * 100); |
| 8 | One divided by the tangent of C | double cotC = 1.0 / tan(C); |
| 9 | Absolute value of the Cotangent of C * 1000 Rounded | double I = round(abs(cotC) * 1000); |
| 10 | Absolute value of the Tangent of D | double J = abs(tan(D)); |
| 11 | D / C (Integer Division) | int K = D / C; |

The engineered dataset (10BillionEngineered) was generated using the latent-factor model shown in Table 2. Three latent factors were defined by variables B, C, and D. Variables E and H were generated primarily from the first latent factor, variables F and I from the second latent factor, and variables G and J from the third latent factor. Variable K was generated as a cross-loading variable influenced by both the second and third latent factors. Independent random error terms were added to the derived variables to prevent perfect collinearity.

Table 2 – Variables in "Engineered" 10-Billion Observation dataset

| Col | Description | Formula |
|---|---|---|
| 1 | Random Number B between 3 and 8 | double B = randBbetween(3,8); |
| 2 | Random Number C between 1 and 10 | double C = randBbetween(1,10); |
| 3 | Random Number D between 1 and 100 | double D = randBbetween(1,100); |
| 4 | Variable E (primarily associated with PC1) | double E = 0.85*PC1 + 0.10*PC2 + 0.05*PC3 + epsE; |
| 5 | Variable F (primarily associated with PC2) | double F = 0.10*PC1 + 0.85*PC2 + 0.05*PC3 + epsF; |
| 6 | Variable G (primarily associated with PC3) | double G = 0.05*PC1 + 0.10*PC2 + 0.85*PC3 + epsG; |
| 7 | Variable H (primarily associated with PC1) | double H = 0.90*PC1 + 0.05*PC2 + 0.05*PC3 + epsH; |
| 8 | Variable I (primarily associated with PC2) | double I = 0.05*PC1 + 0.90*PC2 + 0.05*PC3 + epsI; |
| 9 | Variable J (primarily associated with PC3) | double J = 0.05*PC1 + 0.05*PC2 + 0.90*PC3 + epsJ; |
| 10 | Cross-loading variable K | double K = 0.50*PC2 + 0.50*PC3 + epsK; |

*Streaming Computation of Sufficient Statistics*

Because datasets containing 10 billion and 1 trillion observations were impractical to retain in memory, observations were processed in a streaming manner. Rather than storing individual records, only column sums and the cross-product matrix ($X'X$, hereafter denoted XtX) were accumulated. These sufficient statistics were subsequently used to compute means, covariance matrices, correlation matrices, and PCA solutions.

Let $X$ represent the $N \times p$ data matrix where p = 10 variables. The cross-product matrix was computed as:

$$X^t X = X'X$$

For each generated observation vector $xk$, the column sums and cross-product matrix were updated according to:

$$column\ sums \leftarrow column\ sums + xk$$

$$X^t X \leftarrow X^t X + xk\ xk'$$

After all observations had been processed, column means were obtained from the column sums divided by N, and covariance matrices were computed from $X^t X$ and the vector of means. Correlation matrices were then derived from the covariance matrices and used as input for PCA.

The column sums and $X^t X$ matrix were updated continuously as observations were generated. This approach was adopted because retaining datasets containing 10 billion and 1 trillion observations in memory during analysis was impractical within the memory constraints of the GPU-accelerated system used in this study. By accumulating sufficient statistics during data generation, subsequent analyses could be performed without retaining the original observations. The section below demonstrates the calculations that were used. Because these sufficient statistics contain the information required to compute means, covariance matrices, and correlation matrices, individual observations did not need to be retained after updating the running totals. The approach eliminated the need to store the complete dataset while preserving all information required for the computation of means, covariance matrices, correlation matrices, and PCA solutions.

*Covariance and Correlation Matrices*

After data generation was complete, column means were computed as:

$$\mu_i = \Sigma x_i / N$$

Covariance matrices were then calculated from the accumulated sufficient statistics using:

$$Cov(X) = (X^t X / N) - \mu\mu'$$

Correlation matrices were obtained by standardizing the covariance matrices:

$$Corr(i,j) = Cov(i,j) / \sqrt{[Cov(i,i)\ Cov(j,j)]}$$

*Principal Component Analysis*

PCA was performed using the correlation matrices derived from each dataset. Eigenvalue decomposition of the correlation matrix produced eigenvalues, eigenvectors, and component loading matrices. Eigenvalues were sorted in descending order and explained variance percentages were computed for each component.

For the engineered dataset, the resulting loading matrix was examined to determine whether the recovered principal components corresponded to the latent-factor structure used during data generation.

***Results***

*Stability of PCA at Extremely Large Sample Sizes*

The first objective of this study was to determine whether increasing sample size from 10 billion to 1 trillion observations materially affected PCA results when both datasets were generated from the same underlying process. Comparison of the column means indicated that the two random datasets were nearly identical (Table 3), with most differences occurring only in the fifth or sixth decimal place. Similar agreement was observed in the covariance and correlation matrices.

**Table 3 -- Means for each variable in 3 datasets**

| Variable | 10BillionRandom | 1TrillionRandom | 10BillionEngineered |
|---|---|---|---|
| B | 5.50 | 5.50 | 5.50 |
| C | 5.50 | 5.50 | 5.50 |
| D | 50.50 | 50.50 | 50.50 |
| E | 95.50 | 95.50 | 7.75 |
| F | 23072.10 | 23072.20 | 7.75 |
| G | 64.50 | 64.50 | 43.75 |
| H | 63.46 | 63.46 | 7.75 |
| I | 1775.89 | 1775.90 | 7.75 |
| J | 5.64 | 5.64 | 46.00 |
| K | 14.44 | 14.44 | 28.00 |

The PCA results demonstrated the same pattern. As shown in Table 4, the eigenvalues and explained variance percentages for the 10BillionRandom and 1TrillionRandom datasets were virtually identical. In both datasets, the first three principal components explained approximately 63.8% of the total standardized variance. The corresponding loading patterns were also highly similar. These results suggest that increasing the sample size from 10 billion to 1 trillion observations produced negligible changes in the PCA solution.

Table 4 -- Eigenvalues and Explained Variance

| Dataset | PC1 (%) | PC2 (%) | PC3 (%) | Total PC1–PC3 (%) |
|---|---|---|---|---|
| 10B Random | 25.42 | 21.44 | 16.96 | 63.81 |
| 1T Random | 25.42 | 21.44 | 16.96 | 63.81 |
| 10B Engineered | 60.95 | 21.84 | 17.20 | 100.00 |

The similarity between the two random datasets is consistent with the expectation that estimates of means, variances, covariances, and correlations become increasingly stable as sample size increases. In practical terms, the PCA solution appeared to have converged by the time the sample size reached 10 billion observations. In contrast, the engineered dataset produced substantially different results, reflecting the intentional introduction of latent structure during data generation.

*Recovery of Latent Structure*

The second objective of the study was to determine whether PCA could recover a known latent-factor structure embedded within a dataset containing 10 billion observations. Unlike the random datasets, the engineered dataset was constructed from three latent factors corresponding to variables B, C, and D. Variables E and H were generated primarily from the first latent factor, variables F and I from the second latent factor, and variables G and J from the third latent factor. Variable K was designed as a cross-loading variable influenced by the second and third latent factors. The means, covariance matrix, and correlation matrix for the engineered dataset differed substantially from those observed in the random datasets. These differences reflected the latent-factor relationships intentionally introduced during data generation.

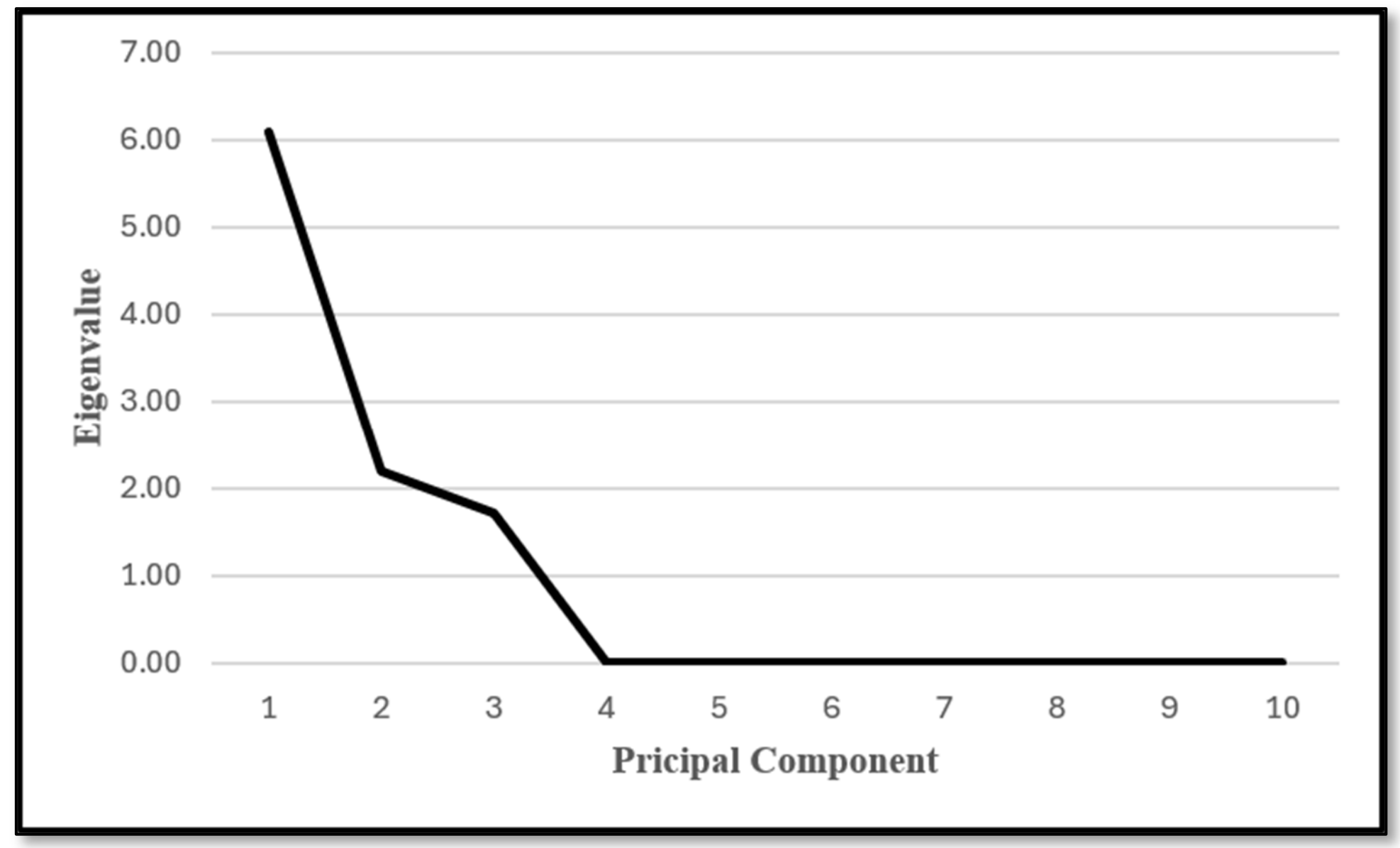

Figure 1 presents the scree plot for the engineered dataset. A pronounced “elbow” was observed after the third principal component. The first three components accounted for 99.996% of the total standardized variance, while the remaining seven components contributed only a negligible amount of additional variance. The corresponding eigenvalues and explained variance percentages are presented in Table 4.

| Table 5 -- Principal Component Loadings (Engineered Dataset) | | | | | | | | | | |
|---|---|---|---|---|---|---|---|---|---|---|
| Variable | Loadings for Principal Components | | | | | | | | | |
| B | 0.25 | -0.26 | -0.93 | 0.00 | 0.00 | 0.01 | 0.00 | 0.00 | 0.00 | 0.00 |
| C | 0.30 | 0.94 | -0.17 | 0.00 | 0.00 | 0.00 | 0.01 | 0.00 | 0.00 | 0.00 |
| D | 0.92 | -0.24 | 0.31 | 0.00 | 0.00 | 0.00 | 0.00 | 0.00 | 0.00 | 0.00 |
| E | 0.86 | -0.22 | -0.46 | -0.01 | 0.00 | 0.00 | 0.00 | 0.00 | 0.00 | 0.00 |
| F | 0.74 | 0.67 | -0.05 | 0.00 | -0.01 | 0.00 | 0.00 | 0.00 | 0.00 | 0.00 |
| G | 0.92 | -0.23 | 0.31 | 0.00 | 0.00 | 0.00 | 0.00 | 0.00 | 0.00 | 0.00 |
| H | 0.83 | -0.29 | -0.48 | 0.01 | 0.00 | 0.00 | 0.00 | 0.00 | 0.00 | 0.00 |
| I | 0.72 | 0.69 | -0.03 | 0.00 | 0.01 | 0.00 | 0.00 | 0.00 | 0.00 | 0.00 |
| J | 0.92 | -0.24 | 0.31 | 0.00 | 0.00 | 0.00 | 0.00 | 0.00 | 0.00 | 0.00 |
| K | 0.94 | -0.15 | 0.29 | 0.00 | 0.00 | 0.00 | 0.00 | 0.00 | 0.00 | 0.00 |

The loading matrix (Table 5) further demonstrated successful recovery of the engineered latent structure. One component was associated primarily with variables D, G, J, and K. A second component was associated primarily with variables C, F, and I. A third component was associated primarily with variables B, E, and H. These loading patterns correspond closely to the latent-factor model used to generate the dataset (Table 6).

| Table 6 -- Designed Loadings vs. Computed Loadings | |
|---|---|
| Designed | Computed |
| PC1 | PC1 |
| B | D |
| E | G |
| H | J |
| | K |
| | |
| PC2 | PC2 |
| C | C |
| F | F |
| I | I |
| | |
| PC3 | PC3 |
| D | B |
| G | E |
| J | H |
| | |
| Mixed | |
| K | |

*Behavior of the Cross-Loading Variable*

Variable K was designed as a cross-loading variable influenced by the second and third latent factors. PCA associated K primarily with the first recovered component, producing a loading of approximately 0.944 on that component, compared with substantially smaller loadings on the remaining components. Examination of the squared loadings indicated that the majority of the variance represented by K was associated with the first recovered component. Although this result differed from the anticipated balanced loading pattern, it nevertheless demonstrates that PCA successfully incorporated K into the dominant latent structure present in the data.

*Summary of Findings*

Two principal findings emerged from the analyses. First, the PCA solutions obtained from the 10BillionRandom and 1TrillionRandom datasets were nearly identical, indicating substantial stability of PCA at extremely large sample sizes. Second, the engineered dataset produced three dominant principal components that accounted for 99.996% of the total standardized variance and closely matched the latent-factor structure used during data generation. Together, these results demonstrate both the stability of PCA at extreme scales and its ability to recover known latent structure in datasets containing billions of observations.

***Discussion***

*PCA Stability at Extreme Sample Sizes*

The first major finding of this study was the remarkable similarity between the PCA solutions obtained from the 10BillionRandom and 1TrillionRandom datasets. Despite a 100-fold increase in the number of observations, the resulting eigenvalues, explained variance percentages, and loading structures were nearly identical. This result is consistent with classical statistical theory, which suggests that estimates of population characteristics become increasingly stable as sample size increases.

One interpretation of these findings is that PCA had already reached practical convergence by approximately 10 billion observations for the data-generating process examined in this study. While theoretical discussions of convergence often consider the limiting behavior of estimators as sample size approaches infinity, the present study provides an empirical demonstration of that process at a scale rarely examined in applied statistical research. The results suggest that further increases in sample size beyond 10 billion observations may provide little additional information for PCA when the underlying data-generating mechanism remains unchanged.

Although the datasets analyzed in this study were generated synthetically, the findings may have implications for domains in which extremely large datasets are common. Examples include remote sensing, digital elevation modeling, environmental monitoring, sensor networks, and other applications in which observations may number in the millions or billions. If the underlying covariance structure remains stable, the present results suggest that PCA solutions may reach practical convergence well before all available observations have been incorporated into the analysis.

*Recovery of Latent Structure*

The second major finding was the successful recovery of the engineered latent-factor structure. The engineered dataset was intentionally designed to contain three dominant factors, and PCA identified three dominant principal components that together explained 99.996% of the total standardized variance. The loading patterns recovered by PCA corresponded closely to the latent-factor model used during data generation. Variables associated with each latent factor clustered together in the recovered component structure, indicating that the underlying relationships embedded within the data remained detectable despite the extreme scale of the dataset.

These findings demonstrate that PCA remains effective as a tool for latent-structure discovery even when datasets contain billions of observations. The ability of PCA to recover meaningful structure does not appear to be diminished by extreme sample size. If anything, the reduction of sampling variability associated with very large datasets may allow latent structure to emerge more clearly. This finding supports the continued use of PCA as an exploratory and descriptive tool in applications where datasets contain millions or billions of observations.

*Behavior of the Cross-Loading Variable*

Variable K was designed as a cross-loading variable influenced by multiple latent factors. Rather than exhibiting a balanced loading pattern across the recovered components, K was associated primarily with the dominant component identified by PCA. This finding suggests that when one latent dimension accounts for substantially more variance than competing dimensions, PCA may preferentially associate cross-loading variables with the dominant factor. Although the loading pattern differed from the original design expectations, the result remained consistent with the overall latent structure embedded within the dataset.

*Implications for Large-N Statistical Analysis*

The findings of this study suggest that classical multivariate procedures remain both stable and informative at scales measured in billions or trillions of observations. An important implication of these findings is that increasing sample size does not necessarily produce meaningful changes in multivariate results once a sufficiently large number of observations has been obtained. For the data-generating process examined in this study, PCA solutions derived from 10 billion observations were nearly indistinguishable from those derived from 1 trillion observations. This suggests that, in some applications, computational effort devoted to acquiring and processing additional observations may yield diminishing returns once practical convergence has been achieved.

Although the streaming approach was originally adopted to overcome practical limitations associated with generating and storing extremely large datasets, the approach has broader applicability. Many scientific and geospatial datasets routinely contain millions or billions of observations. Examples include remote sensing products, digital elevation models, environmental monitoring systems, sensor networks, and large-scale financial and commercial databases. In such settings, accumulation of sufficient statistics may provide a practical alternative to retaining complete datasets while still supporting multivariate analyses such as PCA.

Furthermore, the use of streaming sufficient statistics demonstrates that PCA can be performed without retaining the original observations. Only column sums and the XtX matrix were required to compute means, covariance matrices, correlation matrices, and principal components. This approach substantially reduces storage requirements while preserving the information necessary for multivariate analysis. For organizations that routinely manage large datasets, the ability to derive multivariate statistics from sufficient statistics rather than complete data records may offer substantial advantages in terms of storage efficiency and computational scalability.

Although the present study relied on synthetic data, the underlying questions are relevant to many applied domains. Future research should examine whether similar convergence behavior is observed in real-world datasets containing more complex covariance structures. The present study examined PCA because it is one of the most widely used multivariate techniques; however, the broader question concerns the behavior of classical statistical procedures as sample size becomes extraordinarily large. Whether similar patterns of practical convergence occur for other multivariate methods remains an important area for future investigation.

***Conclusion***

This study investigated the stability of Principal Component Analysis (PCA) at extreme sample sizes and its ability to recover known latent structure in very large datasets. By comparing PCA solutions obtained from two randomly generated datasets and one engineered dataset with a known latent-factor structure, the study investigated both the stability of PCA at extreme sample sizes and its ability to recover known latent structure.

The results demonstrated that the PCA solutions obtained from the 10BillionRandom and 1TrillionRandom datasets were nearly identical, suggesting that PCA had reached practical convergence for the data-generating process examined in this study. Despite a 100-fold increase in the number of observations, the resulting eigenvalues, explained variance percentages, and component structures changed only negligibly. These findings provide empirical evidence that classical multivariate procedures can exhibit substantial stability when applied to datasets containing billions or trillions of observations.

The engineered dataset produced three dominant principal components that accounted for 99.996% of the total standardized variance and closely matched the latent-factor structure used during data generation. This result demonstrates that PCA remains effective for recovering meaningful latent structure even when applied to datasets of unprecedented scale.

Finally, the study demonstrated that PCA can be performed using streamed sufficient statistics without retaining the original observations. By accumulating only column sums and the XtX matrix, it was possible to compute means, covariance matrices, correlation matrices, and PCA solutions for datasets containing up to 1 trillion observations. Future research should examine whether similar patterns of practical convergence are observed in real-world datasets and in other multivariate procedures. Such investigations may provide additional insight into the behavior of classical statistical methods as sample size becomes extraordinarily large.